\documentclass[times,review,10pt]{elsarticle}

\usepackage{amssymb}
\usepackage{amsmath}

\usepackage{lineno}

\usepackage{booktabs} 
\usepackage[normalem]{ulem}
\useunder{\uline}{\ul}{}
\usepackage{color}
\usepackage{url}

\journal{Pattern Recognition}

\begin{document}

\begin{frontmatter}



\title{SmoothSync: Dual-Stream Diffusion Transformers for Jitter-Robust Beat-Synchronized Gesture Generation from Quantized Audio}


\author[SIGS]{Yujiao~Jiang\corref{cor1}} 
\ead{jiangyj20@mails.tsinghua.edu.cn}
\cortext[cor1]{Corresponding author}

\author[SIGS]{Qingmin~Liao} 
\author[SIGS]{Zongqing~Lu} 

\affiliation[SIGS]{organization={Shenzhen International Graduate School, Tsinghua University},
            city={Shenzhen},
            postcode={518055}, 
            country={China}}

\begin{abstract}
Co-speech gesture generation is a critical area of research aimed at synthesizing speech-synchronized human-like gestures. 
Existing methods often suffer from issues such as rhythmic inconsistency, motion jitter, foot sliding and limited multi-sampling diversity.
In this paper, we present SmoothSync, a novel framework that leverages quantized audio tokens in a novel dual-stream Diffusion Transformer (DiT) architecture to synthesis holistic gestures and enhance sampling variation. 
Specifically, we (1) fuse audio-motion features via complementary transformer streams to achieve superior synchronization, (2) introduce a jitter-suppression loss to improve temporal smoothness, (3) implement probabilistic audio quantization to generate distinct gesture sequences from identical inputs. 
To reliably evaluate beat synchronization under jitter, we introduce Smooth-BC, a robust variant of the beat consistency metric less sensitive to motion noise. 
Comprehensive experiments on the BEAT2 and SHOW datasets demonstrate SmoothSync's superiority, outperforming state-of-the-art methods by -30.6\% FGD, 10.3\% Smooth-BC, and 8.4\% Diversity on BEAT2, while reducing jitter and foot sliding by -62.9\% and -17.1\% respectively.
The code will be released to facilitate future research.

\end{abstract}




\end{frontmatter}



\section{Introduction}
\label{sec:intro}

\begin{figure}[htbp]
	\centering
	\includegraphics[width=1.0\textwidth]{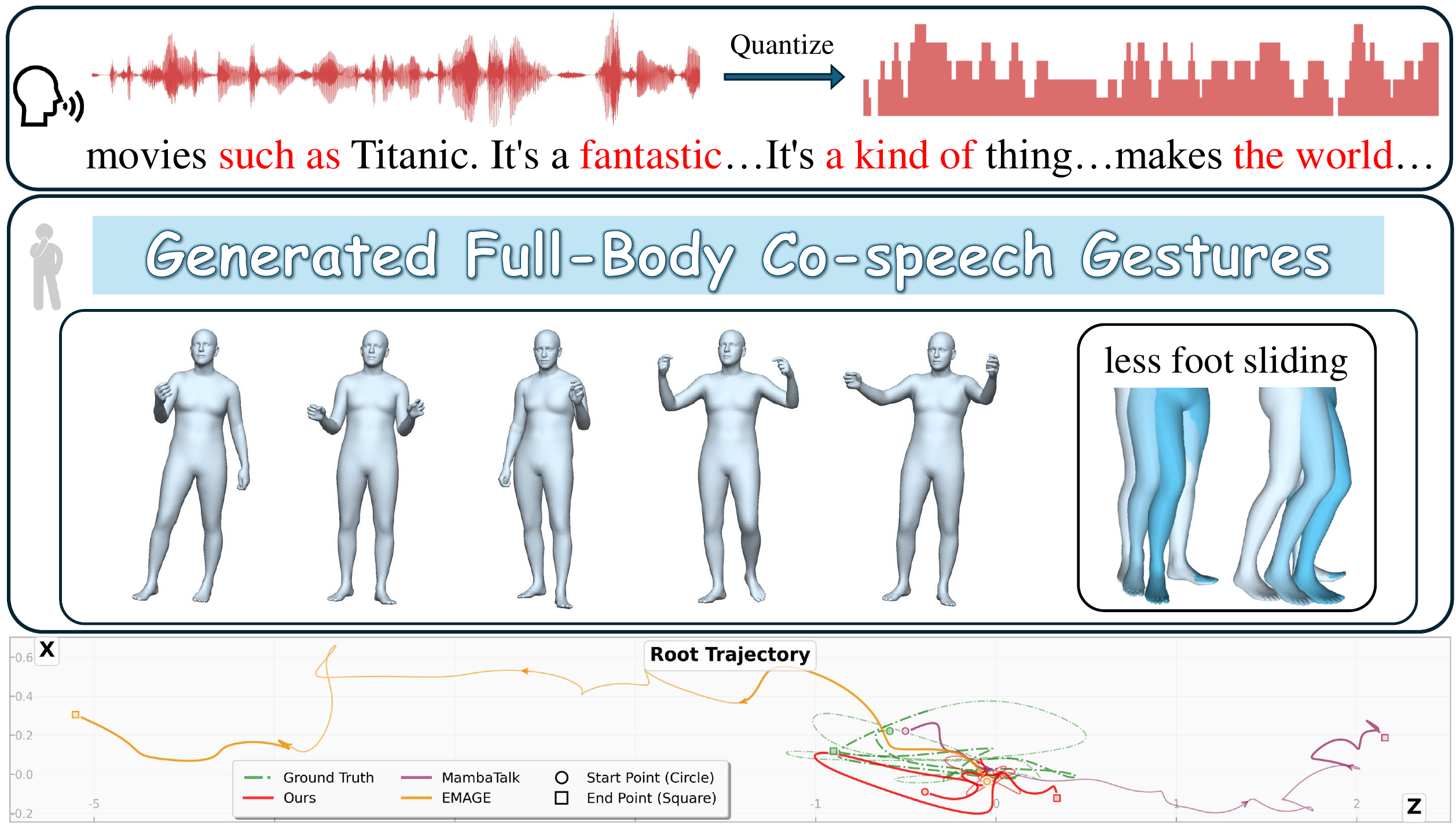}
	\caption{\textbf{SmoothSync} uses quantized audio mel-energy as input and efficiently fuses audio and motion features through a dual-stream network for denoising generation. Our method produces smoother motions with reduced foot sliding and global movement ranges that better match real data. In contrast, previous methods generate motions with severe drift and excessive jittering. \textbf{Zoom in} for better view.}
    \label{fig:teaser}
\end{figure}

Co-speech gesture generation is a fundamental challenge in computer vision, computer graphics and human-computer interaction, with applications ranging from virtual avatars to embodied AI systems. 
The goal is to synthesize natural, expressive body movements that are temporally synchronized with speech audio while maintaining semantic coherence with the spoken content.
Yet, because audio and motion defy a simple one-to-one mapping, the field still struggles to deliver motions that are simultaneously \textbf{rhythmically precise}, \textbf{visually smooth}, and \textbf{stylistically diverse}.

Recent advances have explored various deep learning approaches including Generative Adversarial Networks (GANs)~\cite{ginosar2019learning,habibie2021learning}, Vector Quantized Variational Autoencoder (VQ-VAE)-based autoregressive models~\cite{yi2023generating,liu2024emage,xu2024mambatalk}, diffusion models~\cite{zhu2023taming,alexanderson2023listen}, and state space models~\cite{xu2024mambatalk,fu2024mambagesture}. However, current methods face three critical limitations that hinder practical deployment:
\noindent\textbf{Motion Quality Issues:} Existing approaches suffer from motion artifacts including jitter, foot sliding, and temporal inconsistencies. Many VQ-VAE-based methods~\cite{liu2024emage,xu2024mambatalk,zhang2025echomask} rely on discrete multi-part motion representations that lose fine-grained motion details and struggle to maintain smooth temporal transitions.
\noindent\textbf{Limited Diversity:} Most methods employ large-scale pretrained audio encoders~\cite{whisper,wav2vec2} that produce deterministic outputs with little variation across multiple sampling runs, contradicting natural human behavior where identical speech can be accompanied by different gestures.
\noindent\textbf{Incomplete Body Coverage:} Many approaches~\cite{zhu2023taming,zhi2023livelyspeaker} focus solely on upper-body gestures, neglecting full-body dynamics including global translation, which limits their applicability in immersive environments.

To address these challenges, we propose \textbf{SmoothSync}, a novel dual-stream diffusion transformer framework that revolutionizes how audio and motion features are processed and fused for high-quality gesture synthesis.
Our key insight is that effective audio-motion fusion requires \textit{modality-specific processing} followed by \textit{cross-modal integration}. Unlike existing methods that either process modalities independently or naively concatenate features, our dual-stream architecture maintains separate pathways for audio and motion tokens, allowing each modality to be processed with inductive biases and enabling cross-modal interactions through joint attention mechanisms.

Our approach introduces three key innovations: (1) a \textbf{dual-stream architecture} that processes audio and motion through parallel streams with modality-specific normalization and attention, followed by joint cross-modal attention; (2) a \textbf{jitter-suppress\-ion loss} that explicitly penalizes high-frequency motion artifacts while preserving natural expressiveness; (3) \textbf{quantized mel-spectrogram features} that enable diverse gesture generation for identical inputs while maintaining strong synchronization.

To enable reliable evaluation of rhythmic alignment in the presence of motion artifacts, we introduce the \textbf{Smooth-BC metric}, which filters out spurious beat detections caused by jitter while accurately measuring true rhythmic synchronization. 
Our comprehensive evaluation also includes new metrics for motion quality assessment: Jitter and Foot-Sliding metrics~\cite{shen2024world} that quantify motion artifacts often overlooked by traditional measures, and Inter-Diversity metric to measure the variation between multiple samples generated from the same input.

Experiments on the BEAT2 and SHOW datasets~\cite{liu2024emage,yi2023generating} demonstrate that SmoothSync achieves state-of-the-art performance, generating high-quality, diverse, and rhythmically synchronized full-body gestures with significantly reduced motion artifacts, as shown in Fig.~\ref{fig:teaser}.

Our key contributions are:
\begin{itemize}
    \item We propose \textbf{SmoothSync}, a unified dual-stream diffusion framework that generates high-quality, diverse, and full-body gestures from audio input, achieving substantially lower jitter and superior rhythmic alignment with significant improvements compared to state-of-the-art methods on the BEAT2 and SHOW datasets.
    \item We introduce a \textbf{dual-stream transformer architecture} with modality-specific processing and joint cross-modal attention, coupled with a \textbf{quantized mel-spec\-trogram feature extraction} pipeline that enables diverse gesture generation while maintaining strong synchronization.
    \item We propose \textbf{Smooth-BC}, a robust beat-consistency metric that provides reliable rhythmic assessment by filtering out jitter-induced artifacts, addressing limitations of traditional evaluation protocols.
\end{itemize}

\section{Related Work}
\label{sec:relatedwork}

\noindent\textbf{Co-Speech Gesture Generation.} Early approaches relied on rule-based systems~\cite{cassell1994animated,cassell2001beat,kopp2004synthesizing} and statistical models~\cite{kipp2007towards,levine2010gesture}, which lacked flexibility for natural interaction. 

Deep learning has revolutionized this field through different paradigms.
\textbf{VQ-VAE autoregressive models} like TalkSHOW~\cite{yi2023generating}, EMAGE~\cite{liu2024emage}, and MambaTalk~\cite{xu2024mambatalk} discretize motion into tokens for autoregressive generation, while SemTalk~\cite{zhang2024semtalk} and Echo\-Mask~\cite{zhang2025echomask} employ residual vector quantization (RVQ-VAE). However, discretization errors lead to motion jitter and foot sliding artifacts. These methods divide the body into separate parts, reconstructing each independently, resulting in unnatural and temporally inconsistent gestures.
\textbf{Diffusion models} have emerged as an alternative. DiffGesture~\cite{zhu2023taming}, LivelySpeaker~\cite{zhi2023livelyspeaker}, and recent works like MDT-A2G~\cite{mao2024mdt}, DiffSHEG~\cite{chen2024diffsheg}, and MMoFusion~\cite{wang2025mmofusion} demonstrate stable training and high-quality generation capabilities. Diffusion models handle multi-modal distributions, making them well-suited for gesture generation.
\textbf{State space models} such as MambaTalk~\cite{xu2024mambatalk} and MambaGesture~\cite{fu2024mambagesture} attempt to leverage efficiency for long-sequence modeling but face challenges in capturing dynamic variations across body parts, resulting in jitter artifacts due to error accumulation and insufficient constraints for high-frequency motion details. Related work on motion prediction~\cite{gu2024orientation,dai2023kd} has explored kinematic modeling approaches.

A critical limitation was the focus on upper-body gestures only. Many methods~\cite{liu2022learning,zhu2023taming,yi2023generating,zhi2023livelyspeaker,wu2023audio} neglect full-body dynamics and global translation. The SMPLX~\cite{pavlakos2019expressive} model marked significant advancement, providing unified representation for body, face, and hands, enabling recent works like EMAGE~\cite{liu2024emage} and MambaTalk~\cite{xu2024mambatalk} to tackle full-body synthesis.

EMAGE~\cite{liu2024emage} represents a notable milestone, employing masked audio gesture modeling with four compositional VQ-VAEs for different body parts. While achieving impressive results, it suffers from discrete representation bottlenecks, requires complex multi-stage pipelines, and has limited temporal modeling capabilities. Our approach addresses these issues by directly modeling continuous SMPLX parameters using diffusion.

\noindent\textbf{Audio Feature Design.} Many recent methods~\cite
{chen2024diffsheg,zhang2024semtalk,cheng2025didiffges} employ large-scale pretrained encoders like Wav2vec2~\cite{wav2vec2}, Hubert~\cite{hsu2021hubert}, or Whisper~\cite{whisper}. While providing rich semantic representations, they lead to deterministic outputs with limited inter-sample diversity. In contrast, our method uses quantized Mel-spectrogram features~\cite{stevens1937scale} to enable diverse gesture generation.

\noindent\textbf{Motion Quality Assessment.} The commonly used BC (Beat Consistency) metric~\cite{ellis2007beat,li2021ai,li2022danceformer} is highly sensitive to noise, allowing pure noise motions to achieve high scores, making it unreliable. To address this limitation, we propose Smooth-BC, a robust metric that filters out noise interference for reliable rhythmic alignment assessment.

\section{Method}
\begin{figure*}[t]
	\centering
	\includegraphics[width=\textwidth]{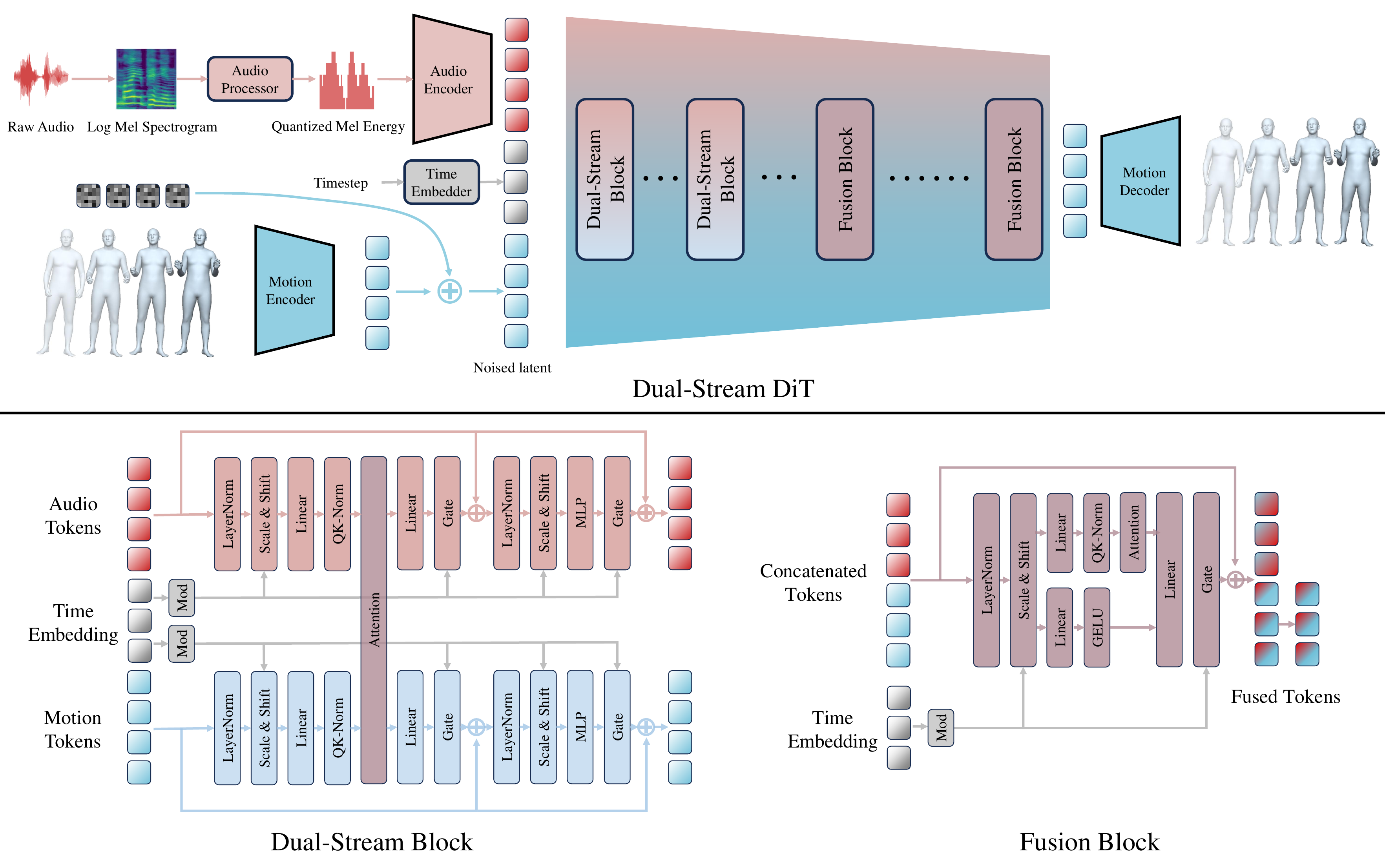}
	\caption{Overview of our Dual-Stream DiT architecture. The framework processes audio and motion through separate pathways before fusing them via specialized transformer blocks, enabling high-quality speech-driven motion generation.}
    \label{fig:pipeline}
\end{figure*}

\subsection{Overall Architecture}

We propose a novel Dual-Stream Diffusion Transformer (DiT) architecture that effectively captures the complex interactions between speech audio and human motion, inspired by recent advances in scalable diffusion transformers~\cite{esser2024scaling}.
As shown in Fig.~\ref{fig:pipeline}, the architecture comprises three key components: (1) an audio encoder that extracts robust multi-modal features from speech signals, (2) a motion encoder that normalizes human motion representations and a motion decoder that transforms processed features back to the original motion space, (3) a dual-stream transformer network with specialized fusion mechanisms.

\subsection{Audio Encoding}
Given input audio, we extract mel-spectrogram features and apply quantization to discretize the audio representations:
\begin{equation}
\mathbf{F}_a = \text{Embed}(\text{Quantize}(\text{MelSpec}(\text{audio}))).
\end{equation}

\noindent\textbf{Audio Feature Extraction:} We extract log mel-spectrogram features from the audio using a 25ms window with 10ms hop length, and apply exponential transformation and compute the magnitude to obtain mel energy.

\noindent\textbf{Mel Energy Augmentation:} 
During training, we apply data augmentation through low-frequency noise augmentation and normalization. We sample random parameters and generate sinusoidal noise:
\begin{equation}
A \sim \mathcal{U}(0, 4), \quad \phi \sim \mathcal{U}(0, 2\pi), \quad f_r \sim \mathcal{U}(150, 1200),
\end{equation}
\begin{equation}
\mathbf{n}(t) = A \sin\left(\frac{2\pi t}{f_r} + \phi\right), \quad t \in \{0, 1, \ldots, T-1\}.
\end{equation}

The augmented features are computed as:
\begin{equation}
\mathbf{x}'_{t,d} = \begin{cases}
\max(x_{\text{thresh}}, (\mathbf{x} + \mathbf{n})_{t,d}) & \text{if } \mathbf{x}_{t,d} > x_{\text{thresh}}, \\
(\mathbf{x} + \mathbf{n})_{t,d} & \text{otherwise},
\end{cases}
\end{equation}
where $x_{\text{thresh}} = \min(\mathbf{x} + 0.1(\max(\mathbf{x}) - \min(\mathbf{x}))$.

Finally, we normalize the features to $[0,1]$:
\begin{equation}
\mathbf{x}_{aug} = \text{clamp}\left(\frac{\mathbf{x}' - \min(\mathbf{x})}{\max(\mathbf{x}) - \min(\mathbf{x})}, 0, 1\right).
\end{equation}

\noindent\textbf{Temporal Quantization:} We apply a three-stage temporal quantization process to audio features for improved temporal alignment and discrete representation.

Given input feature sequence $\mathbf{v} \in [0,1]^{T \times D}$ with length $T$, dimension $D$, and window size $w$, we perform:

\noindent\textbf{Step 1: Temporal Downsampling} with adaptive offset:
\begin{equation}
s = \begin{cases}
\text{randint}(0, w-1) & \text{training} \\
\lfloor w/2 \rfloor & \text{inference}
\end{cases}, \quad \mathbf{v}_{\text{down}} = \mathbf{v}[s::w].
\end{equation}

\noindent\textbf{Step 2: Temporal Upsampling} by repeating each frame $w$ times and adjusting length:
\begin{equation}
\mathbf{v}_{\text{up}} = \text{repeat}(\mathbf{v}_{\text{down}}, w)[:T].
\end{equation}

\noindent\textbf{Step 3: Value Quantization} with adaptive binning:
\begin{equation}
\mathbf{q}_i = \begin{cases}
\mathbb{I}[\mathbf{v}_{\text{up},i} > 0.05] & \text{if } n_{\text{bins}} = 2 \\
\text{clamp}(\lfloor \mathbf{v}_{\text{up},i} \cdot n_{\text{bins}} \rfloor, 0, n_{\text{bins}}-1) & \text{otherwise}
\end{cases}.
\end{equation}

The quantized audio features are then embedded to match the motion feature dimension using a learned embedder.

\subsection{Dual-Stream Diffusion Transformer Network}

\noindent\textbf{Motion Representation:} We represent human motion using SMPLX's~\cite{pavlakos2019expressive} rot6d parameterization. The complete motion vector $\mathbf{m}_t \in \mathbb{R}^{333}$ includes pose rotations $\mathbf{R}_{t} \in \mathbb{R}^{55 \times 6}$ and global translation $\mathbf{T}_{t} \in \mathbb{R}^{3}$.
This representation captures the full complexity of human body motion while maintaining computational efficiency and avoiding the singularities inherent in other rotation representations.


\noindent\textbf{Dual-Stream Architecture:} Our key innovation lies in the dual-stream processing paradigm that maintains separate pathways for audio and motion tokens before strategic fusion. This design allows each modality to be processed with modality-specific inductive biases while enabling cross-modal interactions.

The motion encoder normalizes the motion sequences using precomputed mean and standard deviation statistics:
\begin{equation}
\mathbf{m}_0^{norm} = \frac{\mathbf{m}_0 - \boldsymbol{\mu}_m}{\boldsymbol{\sigma}_m}.
\end{equation}
During training, noise corruption is then applied to the normalized motion:
\begin{equation}
\mathbf{m}_t = \sqrt{\tilde{\alpha}_t} \cdot \mathbf{m}_0^{norm} + \sqrt{1 - \tilde{\alpha}_t} \cdot \boldsymbol{\epsilon},
\end{equation}
where $\mathbf{m}_0^{norm}$ represents the normalized motion, $\boldsymbol{\epsilon}$ is Gaussian noise and $\tilde{\alpha}_t$ follows a predefined noise schedule.

Both audio and motion features are linearly projected to the transformer hidden dimension:
\begin{equation}
\mathbf{X}_a = \text{Linear}_{a}(\mathbf{F}_a), \quad \mathbf{X}_m = \text{Linear}_{m}(\mathbf{m}_t).
\end{equation}

\noindent\textbf{Dual-Stream Blocks:} Each block processes audio and motion tokens through parallel feature extraction followed by joint attention mechanisms. The block consists of:

\begin{itemize}
\item \textbf{Modulation:} Following DiT design principles, we apply adaptive layer normalization conditioned on timestep embeddings for each modality separately:
\begin{equation}
\alpha_{1,i}, \beta_{1,i}, \gamma_{1,i}, \alpha_{2,i}, \beta_{2,i}, \gamma_{2,i} = \text{Mod}_i(\mathbf{t}),
\end{equation}
\begin{equation}
\hat{\mathbf{X}}_i = (1 + \alpha_{1,i}) \cdot \text{LayerNorm}(\mathbf{X}_i) + \beta_{1,i},
\end{equation}
where $i \in \{a, m\}$.

\item \textbf{Parallel QKV Generation:} Each modality independently generates query, key, and value vectors:
\begin{equation}
\mathbf{Q}_i, \mathbf{K}_i, \mathbf{V}_i = \text{Linear}_i(\hat{\mathbf{X}}_i), \quad i \in \{a, m\}.
\end{equation}

\item \textbf{QK-Normalization:} We apply normalization to query and key vectors to stabilize training:
\begin{equation}
\mathbf{Q}_i, \mathbf{K}_i = \text{RMSNorm}(\mathbf{Q}_i, \mathbf{K}_i), \quad i \in \{a, m\}.
\end{equation}

\item \textbf{Joint Cross-Modal Attention:} Audio and motion tokens are fused through concatenated attention:
\begin{equation}
\mathbf{Q}_j, \mathbf{K}_j, \mathbf{V}_j = \text{Concat}([\mathbf{Q}_a, \mathbf{Q}_m], [\mathbf{K}_a, \mathbf{K}_m], [\mathbf{V}_a, \mathbf{V}_m])
\end{equation}
\begin{equation}
\mathbf{A}_a, \mathbf{A}_m = \text{Split}(\text{Attention}(\mathbf{Q}_j, \mathbf{K}_j, \mathbf{V}_j)).
\end{equation}
\end{itemize}

\noindent\textbf{Residual Connections and MLP:} Each modality applies gated residual connections for both attention and MLP layers:
\begin{equation}
\mathbf{X}_i = \mathbf{X}_i + \gamma_{1,i} \cdot \text{Linear}_i(\mathbf{A}_i),
\end{equation}
\begin{equation}
\mathbf{X}_i = \mathbf{X}_i + \gamma_{2,i} \cdot \text{MLP}_i((1 + \alpha_{2,i}) \cdot \text{LayerNorm}(\mathbf{X}_i) + \beta_{2,i}),
\end{equation}
where $i \in \{a, m\}$.

This integrated dual-stream design enables the model to capture both modality-specific features through parallel QKV generation and cross-modal dependencies through joint attention, essential for high-quality speech-driven motion synthesis.

\noindent\textbf{Fusion Blocks:} In addition to the dual-stream blocks, we employ fusion blocks that process concatenated audio-motion features through unified attention and MLP computations. Each fusion block applies:

\begin{itemize}
\item \textbf{Modulated Processing:} The concatenated features are processed with timestep-conditioned normalization:
\begin{equation}
\alpha_f, \beta_f, \gamma_f = \text{Mod}_f(\mathbf{t}),
\end{equation}
\begin{equation}
\mathbf{X}_{mod} = (1 + \alpha_f) \cdot \text{LayerNorm}(\mathbf{X}_{cat}) + \beta_f.
\end{equation}

\item \textbf{Joint QKV and MLP Generation:} A single linear transformation simultaneously generates attention and MLP features:
\begin{equation}
[\mathbf{QKV}, \mathbf{M}] = \text{Split}(\text{Linear}_1(\mathbf{X}_{mod})),
\end{equation}
where $\mathbf{QKV}$ contains the query, key, and value vectors, and $\mathbf{M}$ contains MLP input features.

\item \textbf{Parallel Processing:} Attention and MLP are computed simultaneously:
\begin{equation}
\mathbf{A}_f = \text{Attention}(\text{RMSNorm}(\text{Split}(\mathbf{QKV}))),
\end{equation}
\begin{equation}
\mathbf{M}_{act} = \text{Activation}(\mathbf{M}).
\end{equation}

\item \textbf{Output Integration:} Results are combined with residual connection:
\begin{equation}
\mathbf{X}_{out} = \mathbf{X}_{cat} + \gamma_f \cdot \text{Linear}_2(\text{Concat}([\mathbf{A}_f, \mathbf{M}_{act}]))
\end{equation}
\end{itemize}

This fusion block design efficiently combines cross-modal attention with feed-forward processing in a single unified computation, enabling effective information integration between audio and motion modalities.

\noindent\textbf{Motion Decoder:} The motion decoder is a simple denormalization layer that transforms the processed latent features back to the original motion space. It applies inverse normalization using precomputed statistics:
\begin{equation}
\mathbf{m}_{final} = \mathbf{X}_m \cdot \boldsymbol{\sigma}_m + \boldsymbol{\mu}_m,
\end{equation}
where $\boldsymbol{\mu}_m$ and $\boldsymbol{\sigma}_m$ are the mean and standard deviation statistics computed from the training data, and $\mathbf{X}_m$ represents the processed motion features from the dual-stream transformer.

\subsection{Long Motion Generation}


For extended motion sequences, we adopt a segment-based generation strategy following MimicMotion~\cite{zhang2025mimicmotion}. We divide the sequence into overlapping temporal segments, with each processed independently. Progressive blending in overlapping regions and motion context from previous segments maintain temporal coherence across segments.
This approach enables our method to generate motion sequences of arbitrary length while preserving both local motion quality and global temporal consistency.

\subsection{Training Objective}

Our training objective consists of three main components: rot6d reconstruction loss, translation reconstruction loss, and jitter loss. These losses together ensure that the 
generated motion is accurate, smooth, and physically plausible.

\noindent\textbf{Reconstruction Losses:}
\begin{equation}
\mathcal{L}_{\text{rot6d}} = \frac{1}{\sum_{t=1}^{T} M_{t}} \sum_{t=1}^{T} M_{t} \cdot \frac{1}{D_{\text{rot6d}}} \left\| \mathbf{R}^{\text{pred}}_{t} - \mathbf{R}^{\text{gt}}_{t} \right\|_2^2,
\end{equation}
\begin{equation}
\mathcal{L}_{\text{trans}} = \frac{1}{\sum_{t=1}^{T} M_{t}} \sum_{t=1}^{T} M_{t} \cdot \frac{1}{D_{\text{trans}}} \left\| \mathbf{T}^{\text{pred}}_{t} - \mathbf{T}^{\text{gt}}_{t} \right\|_2^2,
\end{equation}
where $M_t$ is a validity mask for frame $t$, and $D_{\text{rot6d}}$ is the dimension of the rot6d representation, $D_{\text{trans}}$ is the dimension of the translation vector.

\noindent\textbf{Jitter Loss:}
To suppress high-frequency jitter and ensure temporal smoothness, we introduce a jitter loss based on the third-order finite difference (jerk) of the 3D joint positions:
\begin{equation}
\mathcal{L}_{\text{jitter}} = \frac{1}{\sum_{t=1}^{T-3} M^{\text{jitter}}_{t}} \sum_{t=1}^{T-3} M^{\text{jitter}}_{t} \cdot \frac{1}{D_{\text{joints}}} \left\| \mathbf{J}^{\text{pred}}_{t} - \mathbf{J}^{\text{gt}}_{t} \right\|_2^2,
\end{equation}
where $D_{\text{joints}}$ is the number of joints, and the third-order difference is defined as:
\begin{equation}
\mathbf{J}_{t} = \left\| \left(\mathbf{P}_{t+3} - 3\mathbf{P}_{t+2} + 3\mathbf{P}_{t+1} - \mathbf{P}_{t}\right) \cdot \mathrm{fps}^3 \right\|_2,
\end{equation}
where $\mathbf{P}_t$ denotes the 3D joint positions at frame $t$, and $\mathrm{fps}$ is the sampling frequency.

\noindent\textbf{Total Loss:}
The final training objective is a weighted sum of the above losses:
\begin{equation}
\mathcal{L}_{\text{total}} = \lambda_{\text{rot6d}} \mathcal{L}_{\text{rot6d}} + \lambda_{\text{trans}} \mathcal{L}_{\text{trans}} + \lambda_{\text{jitter}} \mathcal{L}_{\text{jitter}},
\end{equation}
where we set $\lambda_{\text{rot6d}} = 1$, $\lambda_{\text{trans}} = 1$, and $\lambda_{\text{jitter}} = 1\times10^{-9}$.

\section{Experiments}
\label{sec:experiments}

\begin{figure}[htbp]
	\centering
	\includegraphics[width=0.8\textwidth]{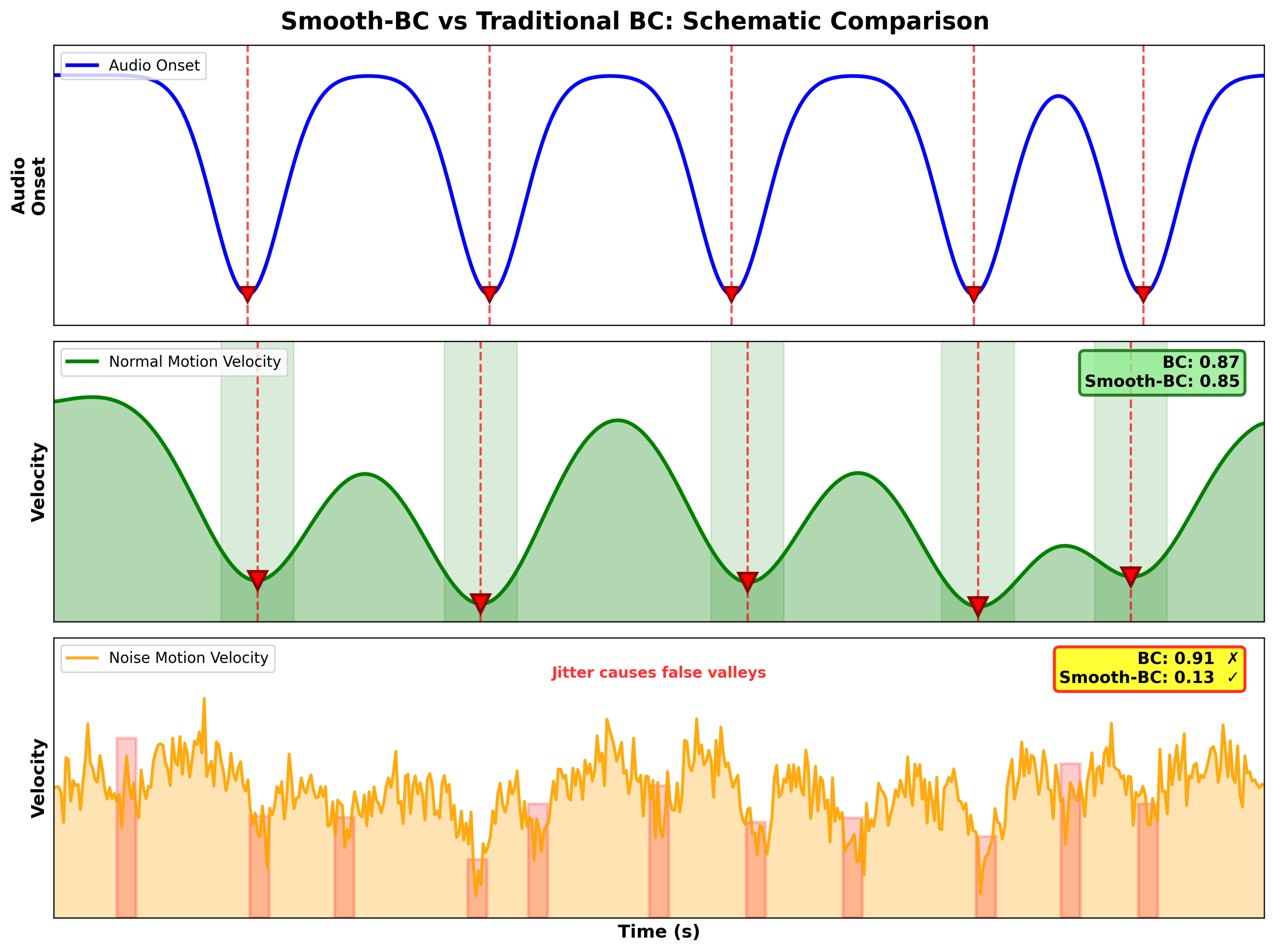}
	\caption{\textbf{Smooth-BC} metric is proposed to exclude motion beats falsely detected due to jitter artifacts by imposing strict constraints on the slope around velocity extrema.}
    \label{fig:bc}
\end{figure}

\subsection{Experimental Setup}

\noindent\textbf{Datasets.} 
We evaluate our method on two datasets: BEAT2 and SHOW.
\textbf{BEAT2}~\cite{liu2024emage} is an extended version of BEAT~\cite{liu2022beat}, containing 76 hours of multi-speaker recordings (30 speakers) with synchronized audio, transcripts, and semantic annotations. The motion data is represented as unified mesh parameters with frame-level labels. Following EMAGE~\cite{liu2024emage}, we adopt the BEAT2-standard subset with an 85\%/7.5\%/7.5\% train/val/test split.
\textbf{SHOW}~\cite{yi2023generating} provides 26.9 hours of real-world talk show videos from four speakers, featuring synchronized 3D body mesh reconstructions and audio streams. Motion is captured at 30 fps using SMPLX~\cite{pavlakos2019expressive} representation, while audio is recorded at 22 kHz sampling rate. We follow the protocol in TalkSHOW~\cite{yi2023generating}, filtering clips longer than three seconds and applying an 80\%/10\%/10\% data split for training, validation, and testing.

\noindent\textbf{Evaluation Metrics.}
We use Fréchet Gesture Distance (FGD)~\cite{yoon2020speech} to measure the realism of generated gestures, Beat Consistency (BC)~\cite{li2021ai} to evaluate audio-motion synchronization, Intra-Diversity~\cite{li2021audio2gestures} to measure the variety within a single generated motion sequence, and Inter-Diversity~\cite{li2021audio2gestures} to measure variety across different sampling runs.
To evaluate motion artifacts such as jitter and foot sliding, we introduce Jitter and Foot-Sliding metrics~\cite{shen2024world}.
Furthermore, we observe that the BC metric is highly sensitive to motion jitter, with pure noise sequences exhibiting abnormally high BC values (BC=7.868, as demonstrated in the Noise Motion variant in Table~\ref{tab:ablation}). To address this critical limitation, we propose the \textbf{Smooth-BC} metric, which imposes strict constraints on the slope around velocity extrema when detecting motion beats, effectively filtering out spurious beat detections caused by jitter artifacts. As illustrated in Fig.~\ref{fig:bc}, traditional BC treats all velocity minima as valid motion beats, allowing high-frequency jitter to artificially inflate scores, while Smooth-BC applies slope constraints to distinguish genuine rhythmic movements from noise-induced fluctuations. The Noise Motion experiment validates this robustness: while pure noise achieves BC=7.868, its Smooth-BC score is nearly zero (0.004965), confirming that Smooth-BC reliably measures true rhythmic synchronization without being misled by motion artifacts.
For comprehensive details on all evaluation metrics, please refer to the supplementary material.

\noindent\textbf{Implementation Details.} 
We conduct training on the BEAT2 dataset using 8 V100 GPUs with a batch size of 16. We employ the AdamW optimizer with a learning rate of 1e-4 and train for 50,000 iterations. 
Each training sequence consists of 320 frames. 
For the diffusion process, we utilize the DDPM schedule with 1,000 steps during training and switch to the DDIM schedule with 50 steps for inference to accelerate generation.
Our method achieves efficient inference at 160 FPS on a single RTX 4090 GPU (5.3$\times$ real-time for 30fps motion), with 154.47M parameters and 2.28 GB memory usage for 320-frame sequences. Detailed computational analysis is provided in the supplementary material.

\begin{table*}[t]
\caption{\textbf{Quantitative comparison on BEAT2 dataset.} SmoothSync achieves the best FGD (motion realism) and Intra-Diversity scores, while maintaining competitive BC performance. The results demonstrate superior motion quality and diversity compared to existing approaches. We report FGD $\times10^{-1}$, BC $\times10^{-1}$. DIV denotes Intra-Diversity.}
\label{tab:beat2_comparison}
\centering
\resizebox{0.5\textwidth}{!}{
    \begin{tabular}{lccc}
    \toprule
    Method & FGD↓ & BC↑ & DIV↑ \\
    \midrule
    S2G~\cite{ginosar2019learning} & 28.15 & 4.683 & 5.971 \\
    Trimodal~\cite{yoon2020speech} & 12.41 & 5.933 & 7.724 \\
    HA2G~\cite{liu2022learning} & 12.32 & 6.779 & 8.626 \\
    DisCo~\cite{liu2022disco} & 9.417 & 6.439 & 9.912 \\
    CaMN~\cite{liu2022beat} & 6.644 & 6.769 & 10.86 \\
    DSG~\cite{yang2023diffusestylegesture} & 8.811 & 7.241 & 11.49 \\
    Habibie \emph{et al.}~\cite{habibie2021learning} & 9.040 & 7.716 & 8.213 \\
    TalkSHOW~\cite{yi2023generating} & 6.209 & 6.947 & {\underline{13.47}} \\
    EMAGE~\cite{liu2024emage} & 5.512 & 7.724 & 13.06 \\
    DiffSHEG~\cite{chen2024diffsheg} & 8.986 & 7.142 & 11.91 \\
    MambaTalk~\cite{xu2024mambatalk} & 5.366 & \textbf{7.812} & 13.05 \\
    EchoMask~\cite{zhang2025echomask} & 4.623 & 7.738 & 13.37 \\
    SemTalk~\cite{zhang2024semtalk} & {\underline{4.278}} & 7.770 & 12.91 \\
    \midrule
    \textbf{SmoothSync} & \textbf{4.151} & {\underline{7.781}} & \textbf{14.00} \\
    \bottomrule
    \end{tabular}
}
\end{table*}

\subsection{Quantitative Results}

\begin{table*}[t]
\caption{\textbf{Extended comparison with motion quality metrics.} SmoothSync significantly outperforms baseline methods across all metrics, achieving the best motion realism, robust beat consistency, Intra and Inter diversity, while substantially reducing motion artifacts. We generate 4 samples using different seeds, and calculate Inter-Diversity between them. We report Smooth-BC $\times10^{-1}$, Jitter $\times10^{2}$, Foot-Sliding $\times10^{-2}$, Inter-Diversity $\times10^{-2}$.}
\label{tab:beat2_jitter}
\centering
\resizebox{\textwidth}{!}{
    \begin{tabular}{lcccccc}
    \toprule
    Method & FGD↓ & Smooth-BC↑ & Intra-Diversity↑ & Jitter↓ & Foot-Sliding↓ & Inter-Diversity↑ \\
    \midrule
    EMAGE~\cite{liu2024emage} & 5.512 & 4.064 & 13.06 & 1.351 & 0.9276 & 0 \\
    MambaTalk~\cite{xu2024mambatalk} & 5.366 & 3.427 & 13.05 & 1.447 & 0.7693 & 0 \\
    \midrule
    \textbf{SmoothSync} & \textbf{3.723} & \textbf{4.484} & \textbf{14.16} & \textbf{0.5008} & \textbf{0.6376} & \textbf{8.090} \\
    \bottomrule
    \end{tabular}
}
\end{table*}

SmoothSync achieves state-of-the-art performance across comprehensive evaluation metrics, demonstrating substantial improvements in motion quality, rhythmic alignment, and diversity generation.

\noindent\textbf{Motion Realism and Diversity.} 
Table~\ref{tab:beat2_comparison} presents our quantitative comparison with baseline methods on the BEAT2 dataset. SmoothSync achieves the \textbf{best FGD score}, indicating superior motion realism. More remarkably, our method achieves the \textbf{highest Intra-Diversity score}, demonstrating that SmoothSync generates motions with richer variation within individual sequences. While maintaining competitive Beat Consistency performance (second-best), our method successfully balances rhythmic alignment with motion diversity—a challenging trade-off that previous methods struggled to achieve.

\noindent\textbf{Motion Quality and Artifacts Reduction.} 
Table~\ref{tab:beat2_jitter} provides extended evaluation focusing on motion quality metrics that are crucial for practical applications. SmoothSync demonstrates exceptional performance across all quality measures: (1) FGD improvement of 30.6\% compared to MambaTalk, indicating substantially more realistic motion generation; (2) Smooth-BC enhancement of 10.3\% over EMAGE, proving superior rhythmic alignment when motion artifacts are properly filtered; (3) \textbf{Jitter reduction of 62.9\%} compared to EMAGE, demonstrating significantly smoother motion generation; (4) Foot-Sliding reduction of 17.1\% compared to MambaTalk, indicating better foot contact consistency.
Most notably, SmoothSync is the \textbf{only method that achieves non-zero Inter-Diversity}, while both EMAGE and MambaTalk score 0, highlighting our method's unique capability to generate diverse gesture variations for identical audio inputs—a critical requirement for natural human-like behavior that existing methods completely fail to address.

\begin{table*}[t]
\caption{\textbf{Quantitative comparison on SHOW dataset.} SmoothSync outperforms TalkSHOW across all evaluation metrics, demonstrating improved motion realism (FGD), robust beat consistency (Smooth-BC), enhanced diversity (Intra-Diversity and Inter-Diversity), and significantly reduced jitter artifacts. The results validate our method's generalization capability to in-the-wild talk show recordings.}
\label{tab:talkshow}
\centering
\resizebox{0.8\textwidth}{!}{
    \begin{tabular}{lcccccc}
    \toprule
    Method & FGD↓ & Smooth-BC↑ & Intra-Diversity↑ & Jitter↓ & Inter-Diversity↑ \\
    \midrule
    TalkSHOW~\cite{yi2023generating} & 5.072 & 2.396 & 3.291 & 0.6795 & 2.295 \\
    \midrule
    \textbf{SmoothSync} & \textbf{4.996} & \textbf{2.749} & \textbf{3.959} & \textbf{0.2987} & \textbf{2.902} \\
    \bottomrule
    \end{tabular}
}
\end{table*}

\begin{figure*}[t]
	\centering
	\includegraphics[width=\textwidth]{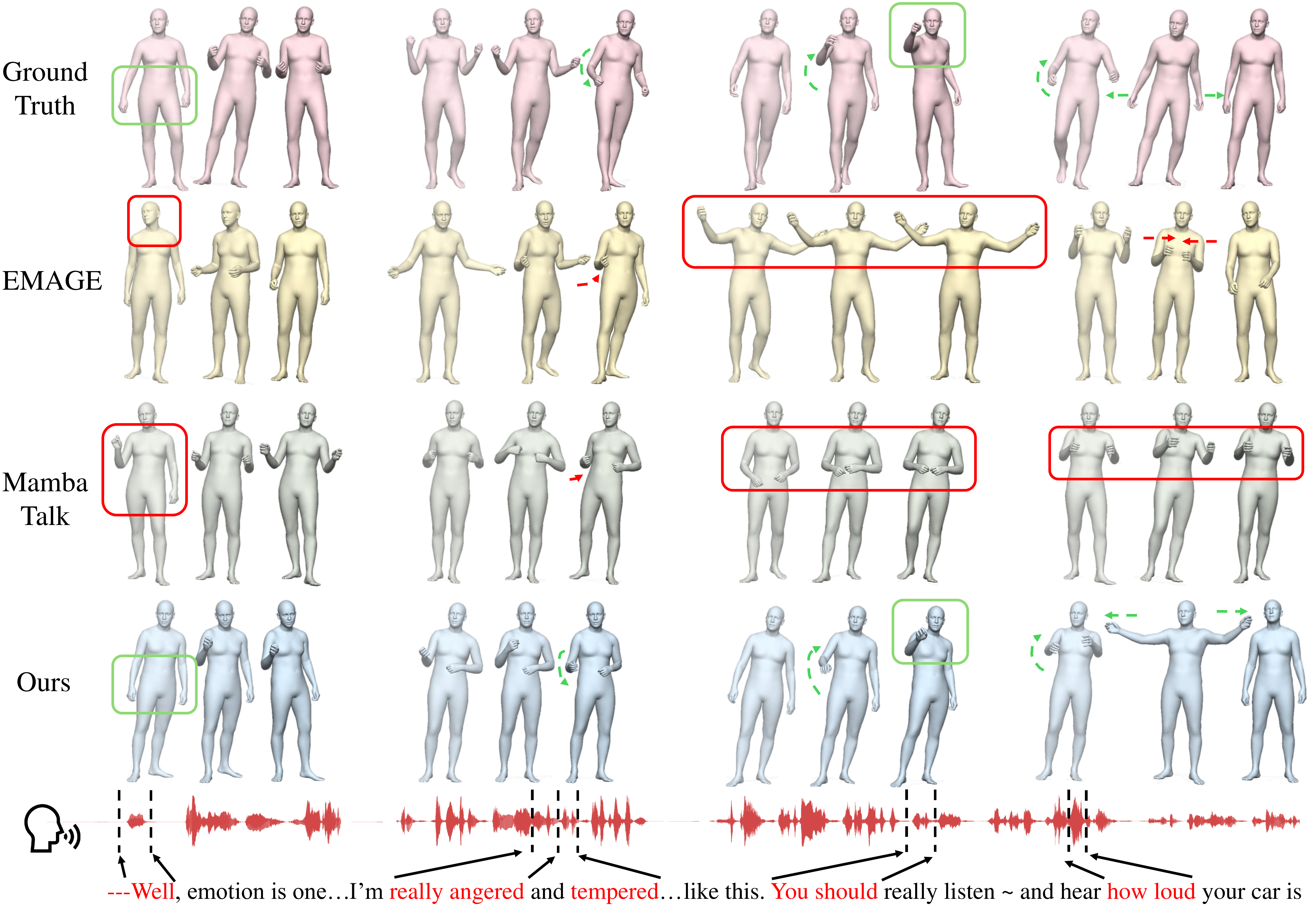}
	\caption{\textbf{Comparison on BEAT2 Dataset}. Compared to baseline methods, our method generates motion that appropriately pauses during speech pauses, performs correct up-and-down hand movements at emphasis points following speech rhythm, and produces gestures closer to ground truth with semantically appropriate hand movements at strong semantic cues like ``You should''. At ``how loud'', SmoothSync generates large-amplitude outward stretching movements with vertical directions that don't completely match ground truth, demonstrating SmoothSync's diversity capability.}
    \label{fig:comparison}
\end{figure*}

\noindent\textbf{Cross-Dataset Generalization.}
To further validate the generalization capability of our approach, we conduct additional experiments on the SHOW dataset~\cite{yi2023generating}, which features in-the-wild talk show recordings with different characteristics compared to BEAT2. Table~\ref{tab:talkshow} presents quantitative comparison with TalkSHOW on the SHOW dataset. SmoothSync demonstrates superior performance \textbf{across all metrics}: 1.5\% FGD improvement, 14.7\% Smooth-BC enhancement, 20.3\% Intra-Diversity increase, and 56.0\% jitter reduction. Most notably, SmoothSync achieves 26.4\% higher Inter-Diversity, demonstrating our method's unique capability to generate varied gesture sequences from identical audio inputs. These results validate that our framework generalizes effectively across diverse datasets with different recording conditions, speaking styles, and motion characteristics.

\begin{figure*}[t]
	\centering
	\includegraphics[width=\textwidth]{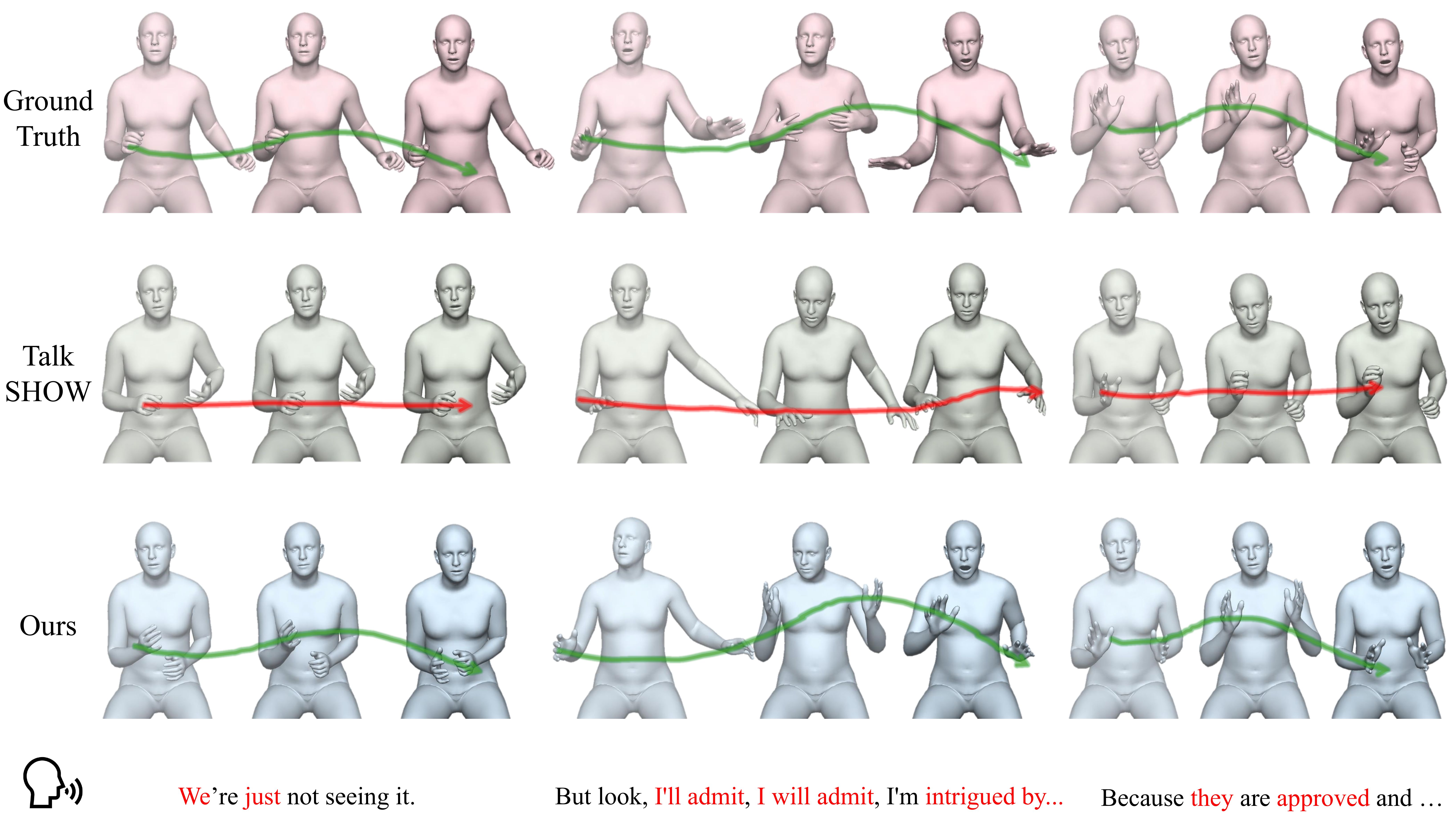}
    \caption{\textbf{Comparison on SHOW Dataset}. Compared to TalkSHOW's limited motion range, our method generates gestures with larger amplitudes and greater diversity, exhibiting rhythmic patterns that align more closely with ground truth. SmoothSync also demonstrates superior semantic matching: at emphasized words such as ``just'' and ``they'', our approach produces gestures that better correspond to the semantic content and prosodic stress of speech.}
    \label{fig:comparison_talkshow}
\end{figure*}

\subsection{Qualitative Results}

Fig.~\ref{fig:comparison} provides comprehensive visual comparison, revealing distinct qualitative advantages of our approach across multiple challenging scenarios.

\noindent\textbf{Rhythmic Precision and Temporal Alignment.} 
The comparison demonstrates SmoothSync's superior ability to capture speech rhythm and temporal dynamics. During speech pauses, our method generates \textbf{appropriate stillness}, closely matching ground truth behavior, while baseline methods exhibit continuous unnecessary movements that break natural speech-gesture synchronization. At emphasis points marked by vocal stress, SmoothSync produces \textbf{precise up-and-down hand movements} that align with speech rhythm, demonstrating our dual-stream architecture's effectiveness in capturing audio-motion correlations.

\noindent\textbf{Semantic Coherence and Gesture Appropriateness.} 
At semantically strong cues like ``You should'', SmoothSync generates gestures that are \textbf{significantly closer to ground truth} with semantically appropriate hand movements, indicating superior understanding of speech-gesture semantic relationships. In contrast, EMAGE produces completely incorrect gestures while MambaTalk generates semantically meaningless movements.

\noindent\textbf{Diversity and Natural Variation.} 
The ``how loud'' segment showcases SmoothSync's unique diversity capability. Our method generates \textbf{large-amplitude outward stretching movements} that maintain similar gesture semantics while exhibiting natural variations in execution details, demonstrating natural human-like variability where identical speech can be accompanied by different but equally appropriate gestures.

\noindent\textbf{Qualitative Comparison on SHOW Dataset.}
The qualitative comparison in Fig.~\ref{fig:comparison_talkshow} reveals distinct advantages of our approach on the SHOW dataset. While TalkSHOW generates gestures with limited motion diversity and restricted amplitude, SmoothSync produces movements with larger amplitudes and richer variations that better match the rhythmic patterns observed in ground truth. Furthermore, our method exhibits superior semantic-prosodic alignment—at emphasized words such as ``just'' and ``they'', SmoothSync generates gestures that appropriately reflect the semantic content and prosodic stress, whereas TalkSHOW fails to capture these nuanced speech-gesture relationships. These qualitative observations, combined with the quantitative results in Table~\ref{tab:talkshow}, confirm our framework's strong generalization capability across diverse datasets and recording conditions. 

\begin{figure*}[t]
	\centering
	\includegraphics[width=\textwidth]{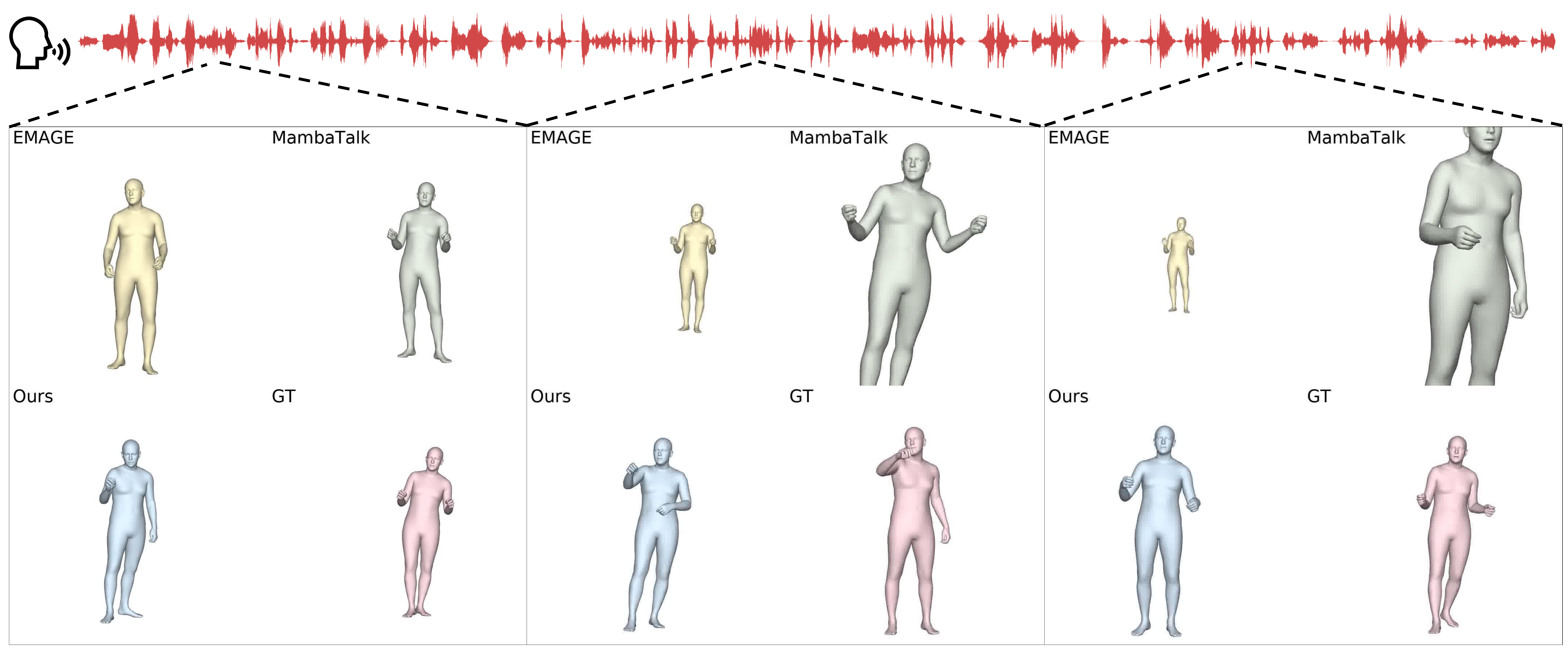}
	\caption{\textbf{Long-Sequence Global Translation Comparison.} Our method demonstrates superior temporal consistency and stability in global translation over extended sequences. While EMAGE and MambaTalk exhibit severe motion drift that progressively accumulates over time, resulting in unrealistic character displacement, SmoothSync maintains translation ranges that closely match ground truth behavior throughout the entire sequence. This comparison highlights our method's effectiveness in preserving long-term spatial coherence while generating natural gestures.}
    \label{fig:transl_comparison}
\end{figure*}

\noindent\textbf{Long-Sequence Consistency and Motion Drift.}
A critical challenge in full-body gesture generation is maintaining temporal consistency and preventing unrealistic motion drift over extended sequences. As illustrated in Fig.~\ref{fig:transl_comparison}, we compare global translation trajectories of generated motions across a long audio sequence. EMAGE and MambaTalk suffer from \textbf{severe motion drift}, with cumulative displacement errors that cause characters to unrealistically slide across the scene. This drift progressively worsens over time, making these methods impractical for long-form content generation. In stark contrast, SmoothSync maintains \textbf{excellent long-sequence consistency}, with global translation ranges that closely align with ground truth behavior throughout the entire sequence. Our method effectively constrains spatial movement within realistic bounds while preserving natural gesture expressiveness, demonstrating the robustness of our dual-stream architecture and jitter-suppression training objective for long-term temporal modeling.

\subsection{Ablation Study}

\begin{table*}[t]
\caption{\textbf{Ablation studies} demonstrate the effectiveness of each component in our framework. The Noise Motion variant uses pure noise as motion input, revealing that the Smooth-BC metric exhibits superior robustness to noise compared to BC metric.}
\label{tab:ablation}
\centering
\resizebox{\textwidth}{!}{
    \begin{tabular}{lccccccc}
    \toprule
    Method & FGD↓ & BC↑ & Smooth-BC↑ & Intra-Diversity↑ & Jitter↓ & Foot-Sliding↓ & Inter-Diversity↑ \\
    \midrule
    SmoothSync & \textbf{3.723} & 7.155 & \textbf{4.484} & \textbf{14.16} & \textbf{0.5008} & \textbf{0.6376} & {\underline{8.090}} \\
    \midrule
    \multicolumn{8}{l}{\textbf{Audio Feature Input (Change from Quantized Mel Energy to: )}} \\
    \midrule
    Mel Energy & 4.798 & 7.236 & \underline{3.753} & 13.10 & \underline{0.9182} & \underline{0.7582} & 7.840 \\
    Whisper~\cite{whisper} & 6.011 & \textbf{7.918} & 0.3294 & 11.61 & 14.20 & 2.471 & 7.235e-4 \\
    Wav2vec2~\cite{wav2vec2} & 8.914 & {\underline{7.858}} & 0.1935 & 9.420 & 9.000 & 1.886 & 5.235e-3 \\
    \midrule
    \multicolumn{8}{l}{\textbf{DiT Achitecture (Change from Dual-Stream to: )}} \\
    \midrule
    Decoder Only & 5.964 & 7.497 & 2.988 & 11.92 & 1.418 & 0.7838 & 6.650 \\
    \midrule
    \textbf{-- jitterloss} & {\underline{4.644}} & 7.796 & 3.244 & {\underline{13.85}} & 1.683 & 1.094 & \textbf{9.069} \\
    \midrule
    \textbf{Noise Motion} & 66.74 & 7.868 & 0.004965 & 13.57 & 78.66 & 17.29 & 1.564 \\
    \bottomrule
    \end{tabular}
}
\end{table*}

To systematically validate the effectiveness of each component in our framework, we conduct comprehensive ablation experiments on the BEAT2 dataset. We analyze the impact of different audio feature representations, network architectures, and training objectives.

\noindent\textbf{Effect of Audio Feature Input.}
To validate the effectiveness of our proposed quantized mel energy, we compare experiments using mel energy, Whisper~\cite{whisper}, and Wav2\-vec2~\cite{wav2vec2} features as input.
As shown in Table~\ref{tab:ablation}, experiments using mel energy show decreased performance in Smooth-BC, Intra-Diversity, and Inter-Diversity metrics compared to quantized mel energy, indicating that the quantization enhancement operation effectively improves model generalization and diversity.
While experiments with Whisper and Wav2vec2 features achieve higher BC scores (7.918, 7.858), they exhibit extremely high jitter metrics and dramatically low Smooth-BC scores (0.3294, 0.1935), revealing that their high BC values are artificially inflated by jitter artifacts rather than true rhythmic alignment. In contrast, our quantized mel-energy approach maintains both strong BC and robust Smooth-BC, demonstrating genuine rhythmic synchronization without motion artifacts. 
Additionally, the particularly low Inter-Diver\-sity scores of large-scale pre-trained models indicate excessive information content that leads to severe overfitting on the BEAT2 dataset and loss of generalization capability.

\noindent\textbf{Effect of DiT Architecture.}
To validate the effectiveness of our proposed Dual-Stream DiT architecture, we compare experiments using Decoder Only and Dual-Stream architectures.
In the Decoder Only architecture, audio conditions are injected into the network through cross-attention, whereas our Dual-Stream architecture performs feature learning on both audio and motion modalities before fusion.
As shown in Table~\ref{tab:ablation}, experiments with the Decoder Only architecture perform inferior to the Dual-Stream architecture across all metrics, indicating that the Dual-Stream architecture better integrates audio and motion information, generating motions with stronger rhythmic sense and more natural movements.

\noindent\textbf{Effect of Jitter Loss.}
We analyze the impact of jitter loss and its weight $\lambda_{\text{jitter}}$ on generation quality. As shown in Table~\ref{tab:ablation}, removing jitter loss results in significantly improved BC metrics while Smooth-BC decreases, and Jitter and Foot-Sliding metrics increase substantially, demonstrating that jitter loss effectively suppresses motion jitter while maintaining rhythmic consistency.

\begin{table}[t]
\caption{\textbf{Jitter loss weight sensitivity.} Although $\lambda_{\text{jitter}}=10^{-9}$ appears small, it operates on jerk magnitude with $\text{fps}^3=27,000$ scaling. $\lambda_{\text{jitter}}=10^{-9}$ achieves optimal balance between realism, diversity, and jitter reduction.}
\label{tab:jitter_weight}
\centering
\resizebox{0.6\textwidth}{!}{
    \begin{tabular}{lcccc}
    \toprule
    $\lambda_{\text{jitter}}$ & FGD↓ & Smooth-BC↑ & Diversity↑ & Jitter↓ \\
    \midrule
    0 (no loss) & 4.644 & 3.244 & 13.85 & 1.683 \\
    $10^{-11}$ & 4.136 & 3.421 & 13.41 & 1.474 \\
    $10^{-10}$ & {\underline{4.113}} & 4.087 & {\underline{14.00}} & 1.076 \\
    $\mathbf{10^{-9}}$ & \textbf{3.723} & {\underline{4.484}} & \textbf{14.16} & {\underline{0.5008}} \\
    $10^{-8}$ & 5.096 & \textbf{4.591} & 13.81 & \textbf{0.3699} \\
    \bottomrule
    \end{tabular}
}
\end{table}

We further conduct sensitivity analysis on $\lambda_{\text{jitter}}$ as shown in Table~\ref{tab:jitter_weight}. Our selected value $\lambda_{\text{jitter}}=10^{-9}$ achieves the best FGD and Diversity while reducing jitter by 70.2\% compared to no jitter loss. Larger weights ($10^{-8}$) over-smooth the motion, degrading realism (FGD $\uparrow$37\%) and diversity, while smaller weights provide insufficient jitter suppression.

\noindent\textbf{Validation of Smooth-BC Metric Robustness.}
To validate the robustness of our proposed Smooth-BC metric against motion artifacts, we conduct a diagnostic experiment using pure noise as motion input while keeping the original audio unchanged. As shown in Table~\ref{tab:ablation} (Noise Motion row), this noise sequence achieves an abnormally high BC score of 7.868—comparable to or even exceeding many legitimate gesture generation methods—demonstrating that traditional BC is severely compromised by high-frequency noise. In stark contrast, the same noise sequence obtains a Smooth-BC score of merely 0.004965 (nearly zero), confirming that our proposed metric successfully filters out spurious beat detections caused by jitter. The dramatic disparity between BC and Smooth-BC scores on noise input (7.868 vs. 0.004965, a 1,585$\times$ difference) validates that Smooth-BC provides reliable rhythmic assessment even in the presence of severe motion artifacts, making it a more trustworthy metric for evaluating gesture generation quality.

\section{Conclusion}
\label{sec:conclusion}

In this paper, we presented SmoothSync, a novel dual-stream diffusion transformer framework for generating high-quality, beat-synchronized co-speech gestures. 
Our approach addresses key limitations in existing gesture generation methods by introducing a dual-stream architecture that effectively captures cross-modal relationships between audio and motion, a jitter-suppression loss that significantly reduces motion artifacts while maintaining natural rhythm, and quantized mel-spectrogram features that enable diverse gesture generation for identical inputs. 
We also proposed the Smooth-BC metric, which provides more reliable assessment of rhythmic alignment by filtering out noise interference, addressing the limitations of traditional evaluation metrics. 
Comprehensive experiments on the BEAT2 and SHOW datasets demonstrate that SmoothSync achieves state-of-the-art performance, generating high-quality, rhythmically synchronized, jitter-free, and diverse co-speech gestures.

\appendix
\section{Evaluation Metrics}

\noindent\textbf{Fréchet Gesture Distance (FGD).}
The FGD metric~\cite{yoon2020speech} quantifies the distributional similarity between authentic and synthesized gesture sequences. This measure computes the Wasserstein-2 distance between feature distributions extracted through a pretrained encoder network, following the methodology established in perceptual evaluation for generative models:
\begin{equation}
\text{FGD}(\mathbf{g}, \hat{\mathbf{g}}) = \|\mu_r - \mu_g\|^2 + \text{Tr}\left(\Sigma_r + \Sigma_g - 2(\Sigma_r\Sigma_g)^{1/2}\right),
\end{equation}
where $\mu_r$, $\Sigma_r$ denote the mean and covariance of latent representations $z_r$ derived from real gesture sequences $\mathbf{g}$, while $\mu_g$, $\Sigma_g$ correspond to statistics from generated motion $\hat{\mathbf{g}}$. We use the pretrained network provided by EMAGE~\cite{liu2024emage} on BEAT2 dataset to extract features.

\noindent\textbf{Intra-Diversity.} 
The diversity metric~\cite{li2021audio2gestures} evaluates gestural variability through pairwise motion analysis, with elevated scores reflecting increased motion richness. This measure computes the mean L1 norm between all gesture pair combinations:
\begin{equation}
\text{Intra-Diversity} = \frac{1}{2N(N-1)} \sum_{l=1}^{N} \sum_{j=1}^{N} \left\|p_l^i - p_j^i\right\|_1,
\end{equation}
where $p_l$ denotes joint coordinates at temporal frame $l$. The diversity assessment spans the complete evaluation dataset. Importantly, global translation components are excluded during joint position computation, ensuring the metric captures exclusively local articulatory variations.

\noindent\textbf{Beat Consistency (BC).}
The BC measure~\cite{li2021ai} quantifies temporal synchronization between gestural rhythms and acoustic prosodic patterns. Enhanced BC values indicate superior audio-motion alignment. Audio beats correspond to speech onset detection, while motion beats derive from velocity minima across upper body articulation (finger joints excluded). The consistency calculation proceeds as:
\begin{equation}
\text{BC} = \frac{1}{g} \sum_{b_g \in g} \exp\left(-\frac{\min_{b_a \in a} \|b_g - b_a\|^2}{2\sigma^2}\right),
\end{equation}
where $g$ and $a$ denote the collections of detected gestural and acoustic beat timestamps, respectively.

\noindent\textbf{Smooth-BC.} Our analysis reveals that conventional BC evaluation suffers from jitter sensitivity, where high-frequency noise artificially inflates synchronization scores. The Smooth-BC metric addresses this vulnerability by incorporating peak detection with slope constraints, effectively discriminating between authentic rhythmic patterns and spurious jitter-induced fluctuations.

The Smooth-BC algorithm initiates with upper body velocity computation:
\begin{equation}
v_t = \|\mathbf{J}_{t+1} - \mathbf{J}_t\|_2,
\end{equation}
where $\mathbf{J}_t$ denotes the 3D joint coordinates at temporal frame $t$.

Subsequently, motion beats are detected as velocity minima using a robust peak detection algorithm with slope constraints. For velocity sequence $\mathbf{v} = [v_1, v_2, \ldots, v_T]$, we first invert the velocity to detect minima as peaks: $\tilde{\mathbf{v}} = -\mathbf{v}$. The peak set (corresponding to velocity minima) is defined as:
\begin{equation}
\mathcal{P} = \{i \mid w \leq i \leq T-w, \text{ satisfying conditions } C_1, C_2, C_3\},
\end{equation}
where $w$ is the local window radius. The three conditions are:
\begin{align}
C_1: &\quad \max_{j=i-w}^{i+w-1} \tilde{v}_j \geq h_{\min} \quad \text{(minimum height)}, \\
C_2: &\quad \tilde{v}_i = \max_{j=i-w}^{i+w-1} \tilde{v}_j \quad \text{(local maximum in inverted velocity)}, \\
C_3: &\quad \tilde{v}_{i-w} \leq \cdots \leq \tilde{v}_{i-1} \leq \tilde{v}_i \geq \tilde{v}_{i+1} \geq \cdots \geq \tilde{v}_{i+w-1} \notag \\
&\quad \text{(slope constraint)},
\end{align}
where $h_{\min}$ is the minimum peak height threshold, $w$ is the local window radius, and $i$ is the candidate peak time index. These constraints ensure that only temporally coherent velocity minima (where motion pauses occur) contribute to beat detection, filtering out high-frequency jitter that would otherwise be detected as spurious beats.

\noindent\textbf{Jitter.} 
To quantify motion jitter artifacts, we compute the third-order finite difference (jerk) of 3D joint positions following~\cite{shen2024world}:
\begin{align}
\mathbf{J}_{t,j}^{\text{jerk}} &= \mathbf{J}_{t+1,j} - 3\mathbf{J}_{t,j} + 3\mathbf{J}_{t-1,j} - \mathbf{J}_{t-2,j}, \\
\text{Jitter} &= \frac{1}{T-3} \sum_{t=3}^{T} \frac{1}{J} \sum_{j=1}^{J} \left\|\mathbf{J}_{t,j}^{\text{jerk}}\right\|_2 \times \text{fps}^3,
\end{align}
where $\mathbf{J}_{t,j}$ represents the 3D position of joint $j$ at frame $t$, $T$ is the total number of frames, $J$ is the number of joints, and $\text{fps}$ is the frame rate. The formula computes third-order differences to measure motion smoothness, with higher jitter values indicating more pronounced temporal discontinuities.

\noindent\textbf{Foot-Sliding.} 
Foot sliding artifacts occur when the feet appear to slide across the ground during contact phases. Following~\cite{shen2024world}, we measure foot sliding by computing the movement error of predicted foot vertices during ground contact periods:
\begin{equation}
\text{Foot-Sliding} = \frac{1}{|\mathcal{C}|} \sum_{t \in \mathcal{C}} \left\|\mathbf{F}^{\text{pred}}_{t+1} - \mathbf{F}^{\text{pred}}_t\right\|_2,
\end{equation}
where $\mathcal{C} = \{t | \|\mathbf{F}^{\text{gt}}_{t+1} - \mathbf{F}^{\text{gt}}_t\|_2 < \text{thr}\}$ is the set of foot contact frames, $\mathbf{F}^{\text{gt}}_t$ and $\mathbf{F}^{\text{pred}}_t$ represent the ground truth and predicted foot vertex positions at frame $t$, respectively. The contact threshold $\text{thr}$ is set to $10^{-2}$. This metric calculates the movement error of predicted feet only when they should remain stationary.

\noindent\textbf{Inter-Diversity.} 
To measure the diversity between multiple samples generated from the same input, we compute the multimodality score that evaluates the pairwise differences between gesture sequences generated using different random seeds:
\begin{equation}
\text{Inter-Diversity} = \frac{1}{N \times \lceil \frac{N}{2} \rceil} \sum_{a=1}^{N} \sum_{b=a+1}^{N} \left\|\hat{\mathbf{M}}_a - \hat{\mathbf{M}}_b\right\|_1,
\end{equation}
where $N$ is the number of motion samples generated from the same audio input using different random seeds, $\hat{\mathbf{M}}_a$ and $\hat{\mathbf{M}}_b$ represent different motion sequences generated for identical audio conditioning, and $\|\cdot\|_1$ denotes the L1 norm computed as the mean absolute difference across all elements. Higher Inter-Diversity scores indicate superior capability to generate varied motions from identical inputs, reflecting the model's ability to avoid mode collapse.

We report FGD $\times10^{-1}$, BC $\times10^{-1}$, Smooth-BC $\times10^{-1}$, Jitter $\times10^{2}$, Foot-Sliding $\times10^{-2}$, Inter-Diversity $\times10^{-2}$.

\section{Computational Efficiency Analysis}

We provide detailed computational metrics to demonstrate the practical efficiency of our method:

\noindent\textbf{Inference Performance.}
Our method achieves efficient inference performance suitable for real-time applications. On a single RTX 4090 GPU with DDIM 50-step sampling, SmoothSync generates motion at 160 FPS (frames per second), enabling real-time generation at 5.3$\times$ real-time speed for 30fps motion sequences. This performance makes our method suitable for interactive applications and live performance scenarios.

\noindent\textbf{Model Complexity.}
The complete SmoothSync model contains 154.47M parameters, with the dual-stream transformer network comprising the majority of parameters. The model size is comparable to other transformer-based gesture generation methods while achieving superior performance.

\noindent\textbf{Memory Requirements.}
For processing 320-frame sequences (approximately 10.67 seconds at 30fps), our method requires 2.28 GB GPU memory during inference. This efficient memory footprint enables deployment on consumer-grade hardware with modest GPU specifications, making the method accessible for practical applications.

\noindent\textbf{Scalability.}
Our efficient architecture supports interactive applications and deployment on consumer-grade hardware. The method can generate extended motion sequences through segment-based generation without significant memory overhead, as each segment is processed independently.

\section{Demo Video and Qualitative Comparison}
We provide comprehensive qualitative comparisons between EMAGE, MambaTalk, and SmoothSync in the supplementary material, including demo videos that showcase the generation quality of each method, in the folder \path{Comparison_Videos_with_Baselines/}.

The comparison videos clearly demonstrate that EMAGE and MambaTalk generate motions with significant jitter artifacts and excessive global translation drift, where the generated characters exhibit unrealistic sliding movements across the scene. In contrast, SmoothSync produces remarkably stable motions with global translation ranges that closely match ground truth behavior.
As shown in Fig.~\ref{fig:transl_comparison} (Section 4.3), the global translation comparison over extended sequences reveals that EMAGE and MambaTalk exhibit progressively accumulating drift over time, while our method maintains translation ranges that closely match ground truth behavior, ensuring realistic character positioning throughout the entire sequence.

Furthermore, our method generates motions that maintain temporal smoothness while exhibiting stronger rhythmic alignment with speech audio. The generated gestures effectively convey rich semantic meaning, demonstrating superior understanding of speech-gesture relationships compared to baseline methods.

Additionally, we provide demonstration videos showcasing multiple diverse gesture samples generated from identical audio inputs, highlighting our method's superior Inter-Diversity capabilities, in the folder \texttt{Inter-Diversity Videos/}. In contrast, EMAGE and MambaTalk produce only deterministic outputs, completely lacking the natural variation exhibited by human speakers.

\section{Dataset Details and Licenses}

\noindent\textbf{BEAT2 Dataset.}
The BEAT2 dataset contains 60 hours of synchronized audio-motion data from 25 speakers (12 female, 13 male). The dataset is divided into BEAT2-standard (27 hours) for training and BEAT2-additional (33 hours) for robustness enhancement.

Following EMAGE~\cite{liu2024emage}, we use an 85\%/7.5\%/7.5\% train/validation/test split on BEAT2-standard. Our experiments focus on speaker-2 data for fair comparison with existing methods.

The dataset includes 1762 sequences with an average length of 65.66 seconds per sequence. Each recording captures natural conversational gestures where participants respond to daily questions.

\noindent\textit{License}: BEAT2 is publicly released under the Apache-2.0 license. We follow all data usage terms and conditions specified by the dataset authors.

\noindent\textbf{SHOW Dataset.}
The SHOW dataset~\cite{yi2023generating} comprises 26.9 hours of talk show recordings from four speakers, captured in natural, uncontrolled settings. Motion sequences are annotated with SMPLX full-body parameters at 30 fps, synchronized with corresponding audio tracks sampled at 22 kHz.

Following the experimental protocol in TalkSHOW~\cite{yi2023generating}, we filter out sequences shorter than three seconds and adopt an 80\%/10\%/10\% split for training, validation, and testing. The dataset captures spontaneous conversational gestures in authentic talk show settings, featuring diverse speaking styles. These characteristics make SHOW a valuable benchmark for assessing model robustness and generalization to real-world scenarios.

\noindent\textit{License}: SHOW dataset is available for non-commercial use only. We comply with all licensing terms and use the dataset solely for academic research purposes.

\section{Baseline Reproduction Details}

\noindent\textbf{EMAGE~\cite{liu2024emage}.} We conduct inference using the official implementation and pretrained models provided by the authors. The model uses a VQ-VAE approach with masked audio gesture modeling, employing four separate VQ-VAEs for different body parts (body, hands, face, and global translation). 

\noindent\textbf{MambaTalk~\cite{xu2024mambatalk}.} We perform inference using the official implementation and pretrained models provided by the authors, following the paper's methodology with a Mamba-based architecture for holistic gesture synthesis.

\noindent\textbf{TalkSHOW~\cite{yi2023generating}.} For experiments on the SHOW dataset, we conduct inference using the official implementation and pretrained models provided by the authors. TalkSHOW employs a VQ-VAE-based autoregressive model that generates full-body gestures from audio input.

\noindent\textbf{Evaluation Protocol.} We generate 4 different samples for each test sequence using different random seeds to compute Inter-Diversity metrics. For fair comparison, all methods use the same test set and evaluation scripts. We implement all metrics using identical joint sets and coordinate systems to ensure consistent evaluation.

 \bibliographystyle{elsarticle-num} 
 \bibliography{sample-bibliography}

@String{Computing = "Computing" }

@String{Computer = "{IEEE} Computer" }

@String{Springer = "Springer-Verlag" }

@article{kopp2004synthesizing,
  title={Synthesizing multimodal utterances for conversational agents},
  author={Kopp, Stefan and Wachsmuth, Ipke},
  journal={Computer animation and virtual worlds},
  volume={15},
  number={1},
  pages={39--52},
  year={2004},
  publisher={Wiley Online Library}
}

@inproceedings{cassell1994animated,
  title={Animated conversation: rule-based generation of facial expression, gesture \& spoken intonation for multiple conversational agents},
  author={Cassell, Justine and Pelachaud, Catherine and Badler, Norman and Steedman, Mark and Achorn, Brett and Becket, Tripp and Douville, Brett and Prevost, Scott and Stone, Matthew},
  booktitle={Proceedings of the 21st annual conference on Computer graphics and interactive techniques},
  pages={413--420},
  year={1994}
}

@inproceedings{cassell2001beat,
  title={Beat: the behavior expression animation toolkit},
  author={Cassell, Justine and Vilhj{\'a}lmsson, Hannes H{\"o}gni and Bickmore, Timothy},
  booktitle={Proceedings of the 28th annual conference on Computer graphics and interactive techniques},
  pages={477--486},
  year={2001}
}

@inproceedings{kipp2007towards,
  title={Towards natural gesture synthesis: Evaluating gesture units in a data-driven approach to gesture synthesis},
  author={Kipp, Michael and Neff, Michael and Kipp, Kerstin H and Albrecht, Irene},
  booktitle={International workshop on intelligent virtual agents},
  pages={15--28},
  year={2007},
  organization={Springer}
}

@article{levine2010gesture,
  title={Gesture controllers},
  author={Levine, Sergey and Kr{\"a}henb{\"u}hl, Philipp and Thrun, Sebastian and Koltun, Vladlen},
  journal={ACM Transactions on Graphics},
  volume={29},
  number={4},
  pages={1--11},
  year={2010},
  publisher={Association for Computing Machinery (ACM)}
}

@inproceedings{li2021audio2gestures,
  title={Audio2gestures: Generating diverse gestures from speech audio with conditional variational autoencoders},
  author={Li, Jing and Kang, Di and Pei, Wenjie and Zhe, Xuefei and Zhang, Ying and He, Zhenyu and Bao, Linchao},
  booktitle={Proceedings of the IEEE/CVF International Conference on Computer Vision},
  pages={11293--11302},
  year={2021}
}

@inproceedings{liu2022learning,
  title={Learning hierarchical cross-modal association for co-speech gesture generation},
  author={Liu, Xian and Wu, Qianyi and Zhou, Hang and Xu, Yinghao and Qian, Rui and Lin, Xinyi and Zhou, Xiaowei and Wu, Wayne and Dai, Bo and Zhou, Bolei},
  booktitle={Proceedings of the IEEE/CVF conference on computer vision and pattern recognition},
  pages={10462--10472},
  year={2022}
}

@inproceedings{liu2024emage,
  title={Emage: Towards unified holistic co-speech gesture generation via expressive masked audio gesture modeling},
  author={Liu, Haiyang and Zhu, Zihao and Becherini, Giorgio and Peng, Yichen and Su, Mingyang and Zhou, You and Zhe, Xuefei and Iwamoto, Naoya and Zheng, Bo and Black, Michael J},
  booktitle={Proceedings of the IEEE/CVF Conference on Computer Vision and Pattern Recognition},
  pages={1144--1154},
  year={2024}
}

@article{alexanderson2023listen,
  title={Listen, denoise, action! audio-driven motion synthesis with diffusion models},
  author={Alexanderson, Simon and Nagy, Rajmund and Beskow, Jonas and Henter, Gustav Eje},
  journal={ACM Transactions on Graphics (TOG)},
  volume={42},
  number={4},
  pages={1--20},
  year={2023},
  publisher={ACM New York, NY, USA}
}

@inproceedings{zhu2023taming,
  title={Taming diffusion models for audio-driven co-speech gesture generation},
  author={Zhu, Lingting and Liu, Xian and Liu, Xuanyu and Qian, Rui and Liu, Ziwei and Yu, Lequan},
  booktitle={Proceedings of the IEEE/CVF Conference on Computer Vision and Pattern Recognition},
  pages={10544--10553},
  year={2023}
}

@inproceedings{zhi2023livelyspeaker,
  title={Livelyspeaker: Towards semantic-aware co-speech gesture generation},
  author={Zhi, Yihao and Cun, Xiaodong and Chen, Xuelin and Shen, Xi and Guo, Wen and Huang, Shaoli and Gao, Shenghua},
  booktitle={Proceedings of the IEEE/CVF international conference on computer vision},
  pages={20807--20817},
  year={2023}
}

@inproceedings{xu2024mambatalk,
  title={Mambatalk: Efficient holistic gesture synthesis with selective state space models},
  author={Xu, Zunnan and Lin, Yukang and Han, Haonan and Yang, Sicheng and Li, Ronghui and Zhang, Yachao and Li, Xiu},
  booktitle={The Thirty-eighth Annual Conference on Neural Information Processing Systems},
  year={2024}
}

@inproceedings{fu2024mambagesture,
  title={MambaGesture: Enhancing Co-Speech Gesture Generation with Mamba and Disentangled Multi-Modality Fusion},
  author={Fu, Chencan and Wang, Yabiao and Zhang, Jiangning and Jiang, Zhengkai and Mao, Xiaofeng and Wu, Jiafu and Cao, Weijian and Wang, Chengjie and Ge, Yanhao and Liu, Yong},
  booktitle={Proceedings of the 32nd ACM International Conference on Multimedia},
  pages={10794--10803},
  year={2024}
}

@inproceedings{mao2024mdt,
  title={Mdt-a2g: Exploring masked diffusion transformers for co-speech gesture generation},
  author={Mao, Xiaofeng and Jiang, Zhengkai and Wang, Qilin and Fu, Chencan and Zhang, Jiangning and Wu, Jiafu and Wang, Yabiao and Wang, Chengjie and Li, Wei and Chi, Mingmin},
  booktitle={Proceedings of the 32nd ACM International Conference on Multimedia},
  pages={3266--3274},
  year={2024}
}

@inproceedings{cheng2025didiffges,
  title={DIDiffGes: Decoupled Semi-Implicit Diffusion Models for Real-time Gesture Generation from Speech},
  author={Cheng, Yongkang and Huang, Shaoli and Chen, Xuelin and Ning, Jifeng and Gong, Mingming},
  booktitle={Proceedings of the AAAI Conference on Artificial Intelligence},
  volume={39},
  pages={2464--2472},
  year={2025}
}

@inproceedings{zhang2025echomask,
  title={EchoMask: Speech-Queried Attention-based Mask Modeling for Holistic Co-Speech Motion Generation},
  author={Zhang, Xiangyue and Li, Jianfang and Zhang, Jiaxu and Ren, Jianqiang and Bo, Liefeng and Tu, Zhigang},
  booktitle={Proceedings of the 33rd ACM International Conference on Multimedia},
  pages={10827--10836},
  year={2025}
}

@inproceedings{shen2024world,
  title={World-Grounded Human Motion Recovery via Gravity-View Coordinates},
  author={Shen, Zehong and Pi, Huaijin and Xia, Yan and Cen, Zhi and Peng, Sida and Hu, Zechen and Bao, Hujun and Hu, Ruizhen and Zhou, Xiaowei},
  booktitle={SIGGRAPH Asia 2024 Conference Papers},
  pages={1--11},
  year={2024}
}

@inproceedings{whisper,
  title={Robust speech recognition via large-scale weak supervision},
  author={Radford, Alec and Kim, Jong Wook and Xu, Tao and Brockman, Greg and McLeavey, Christine and Sutskever, Ilya},
  booktitle={International conference on machine learning},
  pages={28492--28518},
  year={2023},
  organization={PMLR}
}

@article{wav2vec2,
  title={wav2vec 2.0: A framework for self-supervised learning of speech representations},
  author={Baevski, Alexei and Zhou, Yuhao and Mohamed, Abdelrahman and Auli, Michael},
  journal={Advances in neural information processing systems},
  volume={33},
  pages={12449--12460},
  year={2020}
}

@article{hsu2021hubert,
  title={Hubert: Self-supervised speech representation learning by masked prediction of hidden units},
  author={Hsu, Wei-Ning and Bolte, Benjamin and Tsai, Yao-Hung Hubert and Lakhotia, Kushal and Salakhutdinov, Ruslan and Mohamed, Abdelrahman},
  journal={IEEE/ACM transactions on audio, speech, and language processing},
  volume={29},
  pages={3451--3460},
  year={2021},
  publisher={IEEE}
}

@article{stevens1937scale,
  title={A scale for the measurement of the psychological magnitude pitch},
  author={Stevens, Stanley Smith and Volkmann, John and Newman, Edwin Broomell},
  journal={The journal of the acoustical society of america},
  volume={8},
  number={3},
  pages={185--190},
  year={1937},
  publisher={Acoustical Society of America}
}

@inproceedings{zhang2025mimicmotion,
  title={MimicMotion: High-Quality Human Motion Video Generation with Confidence-aware Pose Guidance},
  author={Yuang Zhang and Jiaxi Gu and Li-Wen Wang and Han Wang and Junqi Cheng and Yuefeng Zhu and Fangyuan Zou},
  booktitle={International Conference on Machine Learning},
  year={2025}
}

@inproceedings{pavlakos2019expressive,
  title={Expressive body capture: 3d hands, face, and body from a single image},
  author={Pavlakos, Georgios and Choutas, Vasileios and Ghorbani, Nima and Bolkart, Timo and Osman, Ahmed AA and Tzionas, Dimitrios and Black, Michael J},
  booktitle={Proceedings of the IEEE/CVF conference on computer vision and pattern recognition},
  pages={10975--10985},
  year={2019}
}

@inproceedings{esser2024scaling,
  title={Scaling rectified flow transformers for high-resolution image synthesis},
  author={Esser, Patrick and Kulal, Sumith and Blattmann, Andreas and Entezari, Rahim and M{\"u}ller, Jonas and Saini, Harry and Levi, Yam and Lorenz, Dominik and Sauer, Axel and Boesel, Frederic and others},
  booktitle={Forty-first international conference on machine learning},
  year={2024}
}

@article{ellis2007beat,
  title={Beat tracking by dynamic programming},
  author={Ellis, Daniel PW},
  journal={Journal of New Music Research},
  volume={36},
  number={1},
  pages={51--60},
  year={2007},
  publisher={Taylor \& Francis}
}

@inproceedings{li2021ai,
  title={Ai choreographer: Music conditioned 3d dance generation with aist++},
  author={Li, Ruilong and Yang, Shan and Ross, David A and Kanazawa, Angjoo},
  booktitle={Proceedings of the IEEE/CVF international conference on computer vision},
  pages={13401--13412},
  year={2021}
}

@inproceedings{li2022danceformer,
  title={Danceformer: Music conditioned 3d dance generation with parametric motion transformer},
  author={Li, Buyu and Zhao, Yongchi and Zhelun, Shi and Sheng, Lu},
  booktitle={Proceedings of the AAAI Conference on Artificial Intelligence},
  volume={36},
  pages={1272--1279},
  year={2022}
}

@inproceedings{ginosar2019learning,
  title={Learning individual styles of conversational gesture},
  author={Ginosar, Shiry and Bar, Amir and Kohavi, Gefen and Chan, Caroline and Owens, Andrew and Malik, Jitendra},
  booktitle={Proceedings of the IEEE/CVF conference on computer vision and pattern recognition},
  pages={3497--3506},
  year={2019}
}

@article{yoon2020speech,
  title={Speech gesture generation from the trimodal context of text, audio, and speaker identity},
  author={Yoon, Youngwoo and Cha, Bok and Lee, Joo-Haeng and Jang, Minsu and Lee, Jaeyeon and Kim, Jaehong and Lee, Geehyuk},
  journal={ACM Transactions on Graphics (TOG)},
  volume={39},
  number={6},
  pages={1--16},
  year={2020},
  publisher={ACM New York, NY, USA}
}

@inproceedings{liu2022disco,
  title={Disco: Disentangled implicit content and rhythm learning for diverse co-speech gestures synthesis},
  author={Liu, Haiyang and Iwamoto, Naoya and Zhu, Zihao and Li, Zhengqing and Zhou, You and Bozkurt, Elif and Zheng, Bo},
  booktitle={Proceedings of the 30th ACM international conference on multimedia},
  pages={3764--3773},
  year={2022}
}

@inproceedings{liu2022beat,
  title={Beat: A large-scale semantic and emotional multi-modal dataset for conversational gestures synthesis},
  author={Liu, Haiyang and Zhu, Zihao and Iwamoto, Naoya and Peng, Yichen and Li, Zhengqing and Zhou, You and Bozkurt, Elif and Zheng, Bo},
  booktitle={European conference on computer vision},
  pages={612--630},
  year={2022},
  organization={Springer}
}

@inproceedings{yang2023diffusestylegesture,
  title={DiffuseStyleGesture: stylized audio-driven co-speech gesture generation with diffusion models},
  author={Yang, Sicheng and Wu, Zhiyong and Li, Minglei and Zhang, Zhensong and Hao, Lei and Bao, Weihong and Cheng, Ming and Xiao, Long},
  booktitle={Proceedings of the Thirty-Second International Joint Conference on Artificial Intelligence},
  pages={5860--5868},
  year={2023}
}

@inproceedings{habibie2021learning,
  title={Learning speech-driven 3d conversational gestures from video},
  author={Habibie, Ikhsanul and Xu, Weipeng and Mehta, Dushyant and Liu, Lingjie and Seidel, Hans-Peter and Pons-Moll, Gerard and Elgharib, Mohamed and Theobalt, Christian},
  booktitle={Proceedings of the 21st ACM international conference on intelligent virtual agents},
  pages={101--108},
  year={2021}
}

@inproceedings{yi2023generating,
  title={Generating holistic 3d human motion from speech},
  author={Yi, Hongwei and Liang, Hualin and Liu, Yifei and Cao, Qiong and Wen, Yandong and Bolkart, Timo and Tao, Dacheng and Black, Michael J},
  booktitle={Proceedings of the IEEE/CVF Conference on Computer Vision and Pattern Recognition},
  pages={469--480},
  year={2023}
}

@inproceedings{chen2024diffsheg,
  title={Diffsheg: A diffusion-based approach for real-time speech-driven holistic 3d expression and gesture generation},
  author={Chen, Junming and Liu, Yunfei and Wang, Jianan and Zeng, Ailing and Li, Yu and Chen, Qifeng},
  booktitle={Proceedings of the IEEE/CVF Conference on Computer Vision and Pattern Recognition},
  pages={7352--7361},
  year={2024}
}

@inproceedings{zhang2024semtalk,
  title={SemTalk: Holistic Co-speech Motion Generation with Frame-level Semantic Emphasis},
  author={Zhang, Xiangyue and Li, Jianfang and Zhang, Jiaxu and Dang, Ziqiang and Ren, Jianqiang and Bo, Liefeng and Tu, Zhigang},
  booktitle={Proceedings of the IEEE/CVF international conference on computer vision},
  year={2025}
}

@article{wang2025mmofusion,
  title={MMoFusion: Multi-modal co-speech motion generation with diffusion model},
  author={Wang, Sen and Zhang, Jiangning and Tan, Xin and Xie, Zhifeng and Wang, Chengjie and Ma, Lizhuang},
  journal={Pattern Recognition},
  pages={111774},
  year={2025},
  publisher={Elsevier}
}

@article{gu2024orientation,
  title={Orientation-aware leg movement learning for action-driven human motion prediction},
  author={Gu, Chunzhi and Zhang, Chao and Kuriyama, Shigeru},
  journal={Pattern Recognition},
  volume={150},
  pages={110317},
  year={2024},
  publisher={Elsevier}
}

@article{dai2023kd,
  title={KD-Former: Kinematic and dynamic coupled transformer network for 3D human motion prediction},
  author={Dai, Ju and Li, Hao and Zeng, Rui and Bai, Junxuan and Zhou, Feng and Pan, Junjun},
  journal={Pattern Recognition},
  volume={143},
  pages={109806},
  year={2023},
  publisher={Elsevier}
}

@article{wu2023audio,
  title={Audio-driven talking face generation with diverse yet realistic facial animations},
  author={Wu, Rongliang and Yu, Yingchen and Zhan, Fangneng and Zhang, Jiahui and Zhang, Xiaoqin and Lu, Shijian},
  journal={Pattern Recognition},
  volume={144},
  pages={109865},
  year={2023},
  publisher={Elsevier}
}






\end{document}